\documentclass[twocolumn,aps,superscriptaddress,showpacs]{revtex4}

\usepackage{amssymb}
\usepackage{amsmath}
\usepackage{graphicx}
\usepackage[normalem]{ulem}
\usepackage[dvips]{color}

\begin{document}

\title{Pb-Pb collisions at $\sqrt{s_{NN}}=2.76$ TeV in a multiphase transport  model}
\author{Jun Xu}\email{xujun@comp.tamu.edu}
\affiliation{Cyclotron Institute, Texas A\&M University, College
Station, Texas 77843-3366, USA}
\author{Che Ming Ko}\email{ko@comp.tamu.edu}
\affiliation{Cyclotron Institute and Department of Physics and
Astronomy, Texas A\&M University, College Station, Texas 77843-3366, USA}

\date{\today}

\begin{abstract}
The multiplicity and elliptic flow of charged particles produced in
Pb-Pb collisions at center of mass energy $\sqrt{s_{NN}}=2.76$ TeV
from the Large Hadron Collider are studied in a multiphase transport
(AMPT) model. With the standard parameters in the HIJING model,
which is used as initial conditions for subsequent partonic and
hadronic scatterings in the AMPT model, the resulting multiplicity
of final charged particles at mid-pseudorapidity is consistent with
the experimental data measured by the ALICE Collaboration. This
value is, however, increased by about $25\%$ if the
final-state partonic and hadronic scatterings are turned off.
Because of final-state scatterings, particular those among partons,
the final elliptic flow of charged hadrons is also consistent with
the ALICE data if a smaller but more isotropic parton scattering
cross section than previously used in the AMPT model for describing
the charged hadron elliptic flow in heavy ion collisions at the
Relativistic Heavy Ion Collider is used. The resulting transverse
momentum spectra of charged particles as well as the centrality
dependence of their multiplicity density and the elliptic flow are
also in reasonable agreement with the ALICE data. Furthermore, the
multiplicities, transverse momentum spectra and elliptic flows of
identified hadrons such as protons, kaons and pions are predicted.
\end{abstract}


\maketitle

\section{Introduction}
\label{introduction}

Experimental data on Pb-Pb collisions at $\sqrt{s_{NN}}=2.76$ TeV
from the Large Hadron Collider (LHC) by the ALICE Collaboration have
recently become available~\cite{Aam10a,Aam10b,Aam10c,Aam10d}. For
most central collision bins ($0-5\%$), the mid-pseudorapidity
density of charged particles was found to be $1584 \pm 4 ({\rm
stat.}) \pm 76({\rm sys.})$~\cite{Aam10b}, which is a factor of
$2.2$ higher than that observed in central Au-Au collisions at
$\sqrt{s_{NN}}=200$ GeV from the Relativistic Heavy Ion Collider
(RHIC). This multiplicity density was reproduced by the HIJING2.0
model with a more modern set of parton distribution
functions~\cite{Den10a} and has helped to constrain the gluon
shadowing parameter in the model~\cite{Den10b}. Furthermore, the
elliptic flow in non-central collisions was found to have values
similar to those in collisions at RHIC energies~\cite{Aam10a}.
According to Ref.~\cite{Luz10}, the similarity between the elliptic
flows at LHC and RHIC is consistent with the predictions of the
viscous hydrodynamic model. In Ref.~\cite{Lac10}, this similarity
has further led to the conclusion that the specific viscosity
$\eta_s/s$ of the quark-gluon plasma produced in heavy ion
collisions at LHC has a similar value as in heavy ion collisions at
RHIC. The observed elliptic flow at LHC can also be described by a
kinetic model with a short-lived parton stage but strong final-state
hadronic scatterings~\cite{Hum10}. As shown in Ref.~\cite{Lin01},
adding final-state scatterings, which are essential for generating
the observed elliptic flow~\cite{Zhang99}, to the HIJING model as
implemented in a multiphase transport (AMPT) model~\cite{Lin05}
would reduce the predicted charged particle multiplicity at
mid-rapidity as a result of their appreciable contribution to the
longitudinal work~\cite{Zhang00}. It is thus of interest to study
both the multiplicity and elliptic flow in Pb-Pb collisions at
$\sqrt{s_{NN}}=2.76$ TeV by using the AMPT model.

This paper is organized as follows. In Sec.~\ref{ampt}, we briefly
review the AMPT model and discuss its parameters. We then study in
Sec.~\ref{multiplicity} the multiplicity and in Sec.~\ref{v2} the
elliptic flow in Pb-Pb collisions at $\sqrt{s_{NN}}=2.76$ TeV.
Finally, we give some discussions in Sec.~\ref{discussions} and the
conclusions in Sec.~\ref{summary}.

\section{The AMPT model}
\label{ampt}

The AMPT model is a hybrid model with the initial particle
distributions generated by the HIJING model~\cite{Xnw91} of the
version $1.383$. In the version of string melting, which is used in
the present study, hadrons produced from the HIJING model are
converted to their valence quarks and antiquarks, and their
evolution in time and space is then modeled by the ZPC parton
cascade model~\cite{Zha98} with the differential scattering cross
section
\begin{equation}\label{cross}
\frac{d\sigma}{dt} \approx \frac{9\pi\alpha_s^2}{2(t-\mu^2)^2}.
\end{equation}
In the above, $t$ is the standard Mandelstam variable for four
momentum transfer, $\alpha_s$ is the strong coupling constant, and
$\mu$ is the screening mass in the partonic matter. After stopping
scattering, quarks and antiquarks are converted via a spatial
coalescence model to hadrons, which are followed by hadronic
scatterings via the ART model~\cite{Bal95}.

In previous studies of heavy ion collisions at RHIC, some of the
parameters in the HIJING model were varied in order to reproduce the
measured charged particle multiplicity. Specifically, instead of the
values $a=0.5$ and $b=0.9$ GeV$^{-2}$ used in the HIJING model for
the Lund string fragmentation function $f(z) \propto z^{-1} (1-z)^a
\exp (-b~m_{\perp}^2/z)$, where $z$ is the light-cone momentum
fraction of the produced hadron of transverse mass $m_\perp$ with
respect to that of the fragmenting string, the values $a=2.2$ and
$b=0.5$ GeV$^{-2}$ were used in the AMPT model~\cite{Lin01}. Also,
the values $\alpha_s=0.47$ and $\mu=1.8$ or $2.3$ fm$^{-1}$,
corresponding to a total parton scattering cross section of $10$ or
$6$ mb, were used for the parton scattering cross section in the
AMPT model to describe measured elliptic flow~ \cite{Lin02a,Chen04}
and two-pion correlation functions~ \cite{Lin02b}. In the present
study of heavy ion collisions at LHC, we again vary the values of
these parameters to fit measured data. As shown below, a reasonable
description of the measured charged particle multiplicity density at
mid-pseudorapidity is achieved if the default HIJING values of
$a=0.5$ and $b=0.9$ GeV$^{-2}$ are used in the Lund string
fragmentation function. Also, a smaller QCD coupling constant
$\alpha_s=0.33$ and a larger screening mass $\mu=3.2$ fm$^{-1}$,
corresponding to a smaller ($1.5$ mb) but more isotropic parton
scattering cross section, are needed to reproduce the observed
elliptic flow.

\section{Rapidity distributions and transverse momentum spectra}
\label{multiplicity}

\begin{figure}[h]
\centerline{\includegraphics[scale=0.8]{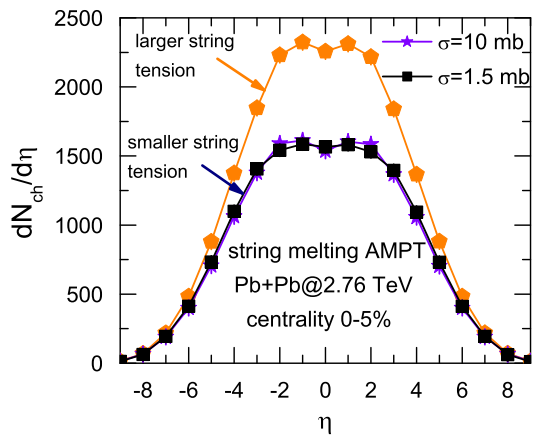}} \caption{(Color
online) Pseudorapidity distribution of charged particles in Pb-Pb
collisions at $\sqrt{s_{NN}}=2.76$ TeV and centrality of $0-5\%$
from the AMPT model with string melting for larger string tension
(filled pentagons) and for smaller string tension with a parton
scattering cross section of $10$ mb (filled stars) or $1.5$ mb
(filled squares).} \label{dndetacom}
\end{figure}

We first show in Fig.~\ref{dndetacom} the pseudorapidity
distribution of charged particles in the most central $5\%$
($\text{b}<3.5$ fm) Pb-Pb collisions at $\sqrt{s_{NN}}=2.76$ TeV
from the AMPT model with string melting using different values for
the Lund string fragmentation parameters and parton scattering cross
sections. It is seen that the values of $a=2.2$ and $b=0.5$
GeV$^{-2}$, that correspond to a larger string tension in the Lund
string fragmentation function, gives a larger multiplicity density
at mid-pseudorapidity than that from the default HIJING values of
$a=0.5$ and $b=0.9$ GeV$^{-2}$ that correspond a smaller string
tension in the fragmentation function. The multiplicity density is,
however, not sensitive to the parton scattering cross section
$\sigma$. For the smaller value of $\sigma=1.5~{\rm mb}$, which is
required to reproduce the measured elliptic flow at LHC as shown
later in Sec.~\ref{v2}, the resulting charged particle multiplicity
of $1,536\pm4$ at mid-pseudorapidity obtained with the default
HIJING parameters for the string fragmentation function is
consistent with the measured data by the ALICE Collaboration. In the
following, we will thus use, unless stated otherwise, those
parameters that correspond to the smaller values of string tension
and parton cross sections in the AMPT model for our study. We note
that in obtaining the relation between the centrality $c$ and the
impact parameter $\text{b}$, we have used the empirical formula
$c=\pi \text{b}^2/\sigma_{\rm in}$~\cite{Bro02} with the
nucleus-nucleus total inelastic cross section $\sigma_{\rm in}
\approx 784$ fm$^2$ calculated from the Glauber model. Also, we have
used in our analysis about $1,000$ AMPT events for the multiplicity
density, transverse momentum spectra, and the total elliptic flow,
and about $10,000$ events for the transverse momentum dependence of
the elliptic flow.

\begin{figure}[h]
\centerline{\includegraphics[scale=0.8]{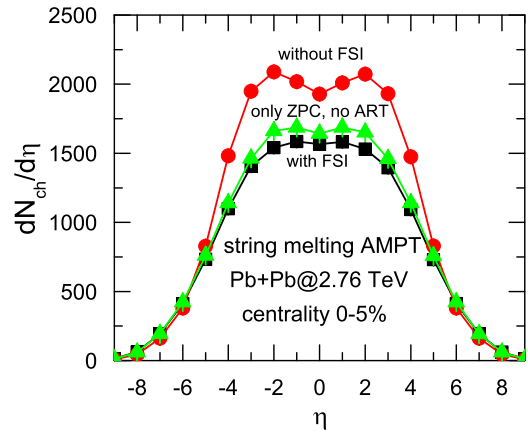}} \caption{(Color
online) Pseudorapidity distribution of charged particles in Pb-Pb
collisions at $\sqrt{s_{NN}}=2.76$ TeV and centrality of $0-5\%$
from the AMPT model with both final-state partonic and hadronic
scatterings (filled squares), with only partonic scatterings (filled
triangles), and without final-state interactions (filled circles).}
\label{dndeta}
\end{figure}

We have also studied the effect of final-state interactions (FSI) on
the charged particle multiplicity. As shown by filled circles in
Fig.~\ref{dndeta}, turning off the final-state partonic and hadronic
scatterings in the AMPT model enhances the charged particle
pseudorapidity distribution. The charged particle multiplicity at
mid-pseudorapidity is now $1925\pm5$, which is essentially the value
from the HIJING1.0 model~\cite{Den10a} and is about
$25\%$ higher than that with the inclusion of
final-state scatterings, which is shown by filled squares.  The
value is reduced to $1642\pm4$ after including partonic scatterings,
and this shows that the charged particle pseudorapidity distribution
is much more sensitive to partonic than hadronic scatterings.

\begin{figure}[h]
\centerline{\includegraphics[scale=0.8]{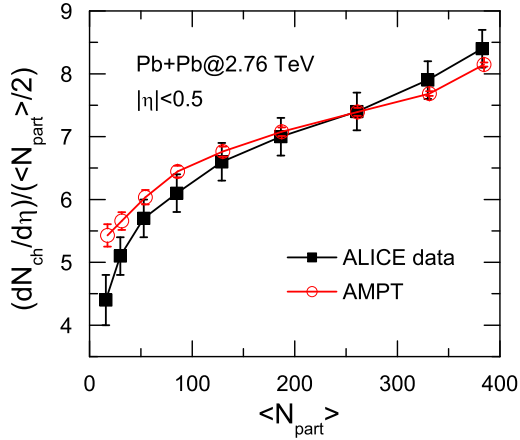}} \caption{(Color
online) Dependence of charged particle pseudorapidity density per
half participant $(dN_{\rm ch}/d\eta)/(\langle N_{\rm
part}\rangle/2)$ on the number of participants in Pb-Pb collisions
at $\sqrt{s_{NN}}=2.76$ TeV from the AMPT model with string melting
and from the ALICE data~\cite{Aam10c}.} \label{dndetacen}
\end{figure}

Figure \ref{dndetacen} displays the dependence of the charged
particle multiplicity density at mid-pseudorapidity ($|\eta|<0.5$)
per 1/2 participant, $(dN_{\rm ch}/d\eta)/(\langle N_{\rm
part}\rangle /2)$, on the number of participants. It is seen that
results from the AMPT model with string melting (open circles) are
roughly consistent with the experimental data, although for peripheral collisions
the values are higher than the ALICE data.

\begin{figure}[h]
\centerline{\includegraphics[scale=0.8]{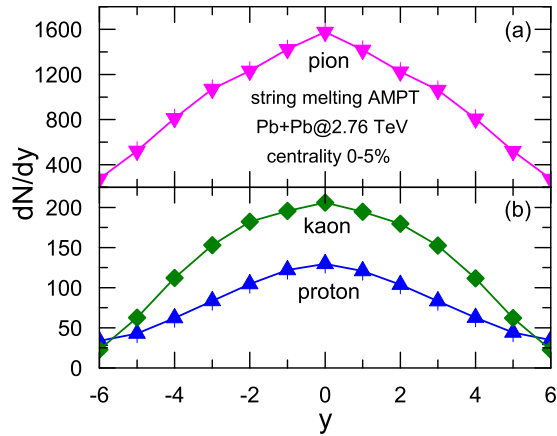}} \caption{(Color
online) Rapidity distributions of protons, kaons and pions in Pb-Pb
collisions at $\sqrt{s_{NN}}=2.76$ TeV and $0-5\%$ centrality from
the AMPT model with string melting.} \label{dndyppk}
\end{figure}

For identified hadrons such as protons, kaons, and pions as well as
their particles and antiparticles, their rapidity distributions in
the $0-5\%$ centrality of same collisions are shown in
Fig.~\ref{dndyppk}. The multiplicity ratio of mid-rapidity protons,
kaons and pions is roughly $1:2:16$. We note that the multiplicities
of each species are again larger if final-state scatterings are not
included.

\begin{figure}[h]
\centerline{\includegraphics[scale=0.8]{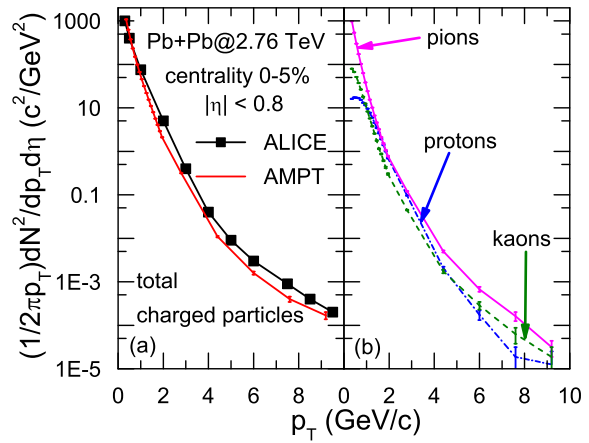}} \caption{(Color
online) Transverse momentum spectrum of mid-pseudorapidity
($|\eta|<0.8$) charged particles (left panel) as well as those of
protons, kaons, and pions (right panel) in Pb-Pb collisions at
$\sqrt{s_{NN}}=2.76$ TeV and $0-5\%$ centrality from the AMPT model
with string melting. Experimental data from the ALICE
Collaboration~\cite{Aam10d} are shown by solid squares in the left
panel.} \label{dndpt}
\end{figure}

In Fig.~\ref{dndpt}, the transverse momentum ($p_T$) spectrum of
mid-pseudorapidity ($|\eta|<0.8$) charged particles (left panel) and
those of protons, kaons and pions (right panel) are shown. It is
seen that the AMPT model describes reasonably the charged particle
$p_T$ spectrum from the ALICE data for small $p_T$ but gives smaller
values for large $p_T$. For identified hadrons, the abundance of
pions is the largest at all $p_T$ except at intermediate $p_T$ where
the proton yield becomes comparable. A similar phenomenon was
observed in heavy ion collisions at RHIC~\cite{Adcox02}, and it was
attributed to the enhanced production of baryons from quark
coalescence~\cite{Hwa03,Greco03,Fries03}.

\section{Elliptic flows}
\label{v2}

\begin{figure}[h]
\centerline{\includegraphics[scale=0.8]{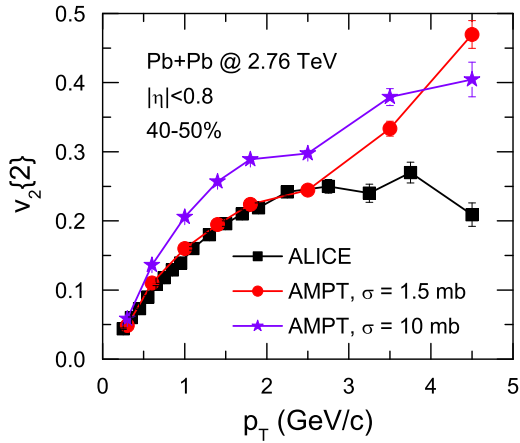}} \caption{(Color
online) Transverse momentum dependence of the elliptic flow obtained
from the two-particle cumulant method for charged particles in Pb-Pb
collisions at $\sqrt{s_{NN}}=2.76$ TeV and $40-50\%$ centrality from
the AMPT model with string melting using a parton scattering cross
section of $1.5$ mb (filled circles) or $10$ mb (filled stars).
Corresponding experimental data from Ref.~{\cite{Aam10a}} are shown
by filled squares.} \label{v2pt}
\end{figure}

In the present study, we determine the elliptic flow using the
two-particle cumulant method~\cite{Wan91,Bor01},
\begin{equation}
v_2\{2\} = \sqrt{\langle \cos(2\Delta\phi) \rangle },
\end{equation}
where $\Delta\phi$ is the azimuthal angular difference between
particle pairs within the same event and $\langle\cdots\rangle$
means average over all possible pairs. The error of the elliptic
flow is calculated by using the method in Ref.~\cite{Bor01}.

In Fig.~\ref{v2pt}, we show by filled circles the transverse
momentum dependence of the elliptic flow $v_2\{2\}$ of
mid-pseudorapidity ($|\eta|<0.8$) charged particles at centrality of
$40-50\%$ ($10.0$ fm $<\text{b}<11.2$ fm) from the AMPT model and
compare it with the ALICE data (filled squares)~\cite{Aam10a}. It is
seen that the elliptic flow from the AMPT model is consistent with
the experimental data at $p_T< 2.5~{\rm GeV}/c$ but is larger at
$p_T>2.5~{\rm GeV}/c$. The latter is likely due to an overestimated
non-flow effect in the AMPT model. We note that a satisfactory
understanding of elliptic flow of high-$p_T$ particles in heavy ion
collisions at RHIC is still lacking~\cite{Adl03b,Shu02}. Using the
larger parton scattering cross section of $10$ mb, corresponding to
a larger strong coupling constant ($\alpha_s=0.47$) and a smaller
screening mass ($\mu=1.8$ fm$^{-1}$) as used in describing the
elliptic flow at RHIC, would lead to a larger elliptic flow (filled
stars) than the ALICE data.

\begin{figure}[h]
\centerline{\includegraphics[scale=0.8]{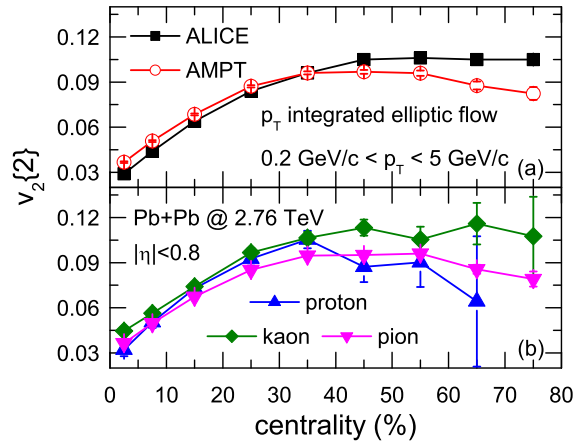}} \caption{(Color
online) Centrality dependence of the elliptic flow obtained from the
two-particle cumulant method for all  charged particles (upper
panel) and for protons, kaons and pions (lower panel) in Pb-Pb
collisions at $\sqrt{s_{NN}}=2.76$ TeV from the AMPT model with
string melting. The ALICE data (filled squares) are from
Ref.~\cite{Aam10a}.} \label{v2cen}
\end{figure}

The centrality dependence of the $v_2\{2\}$ of mid-pseudorapidity
charged particles with transverse momenta $0.2$ GeV/c $<p_T< 5$
GeV/c is shown in the upper panel of Fig.~\ref{v2cen} by using again
the smaller parton scattering cross section of $1.5$ mb. It is seen
that the results from the AMPT model (open circles) are consistent
with the experimental data (filled squares) except for central and
peripheral collisions where they are slightly larger and somewhat
smaller, respectively. The smaller elliptic flow for centralities
larger than $40-50\%$ is likely related to the softer $p_T$
spectrum in the AMPT model. In the lower panel of Fig.~\ref{v2cen},
we compare the centrality dependence of the elliptic flows of
protons, kaons and pions. Interestingly, the elliptic flow is larger
for kaons than for pions.

\section{Discussions}
\label{discussions}

The above results from the AMPT model for heavy ion collisions at
LHC were obtained with different values for the parameters in the
Lund string fragmentation function and in the parton scattering
cross section from those used for heavy ion collisions at RHIC. The
values of $a=2.2$ and $b=0.5$ GeV$^{-2}$ used at RHIC for the Lund
string fragmentation function correspond to a string tension that is
about $7\%$ larger than that for $a=0.5$ and $b=0.9$ GeV$^{-2}$ used
in the present study. A smaller string tension at LHC is consistent
with the fact that the matter formed at LHC is hotter and denser
than that at RHIC. The higher temperature reached at LHC also makes
it possible to understand, according to the lattice
results~\cite{Kac05a,Kac05b}, the smaller QCD coupling constant and
the larger screening mass needed to describe the elliptic flow at
LHC.

We can compare the property of the quark-gluon plasma produced in
heavy ion collisions at LHC with that at RHIC by considering the
ratio of its shear viscosity and entropy density, i.e., the specific
viscosity. In the kinetic theory, the shear viscosity is given by
$\eta_s=4\langle p\rangle/(15\sigma_{\rm tr})$, where $\langle
p\rangle$ is the mean momentum of partons and $\sigma_{\rm tr}$ is
the parton transport or viscosity cross section defined by
\begin{eqnarray}
\sigma_{\rm tr}&=&\int dt\frac{d\sigma}{dt}(1-\cos^2\theta)\nonumber\\
&=&\frac{18\pi\alpha_s^2}{E^2}
\left[\left(1+\frac{2\mu^2}{E^2}\right)\ln\left(\frac{1+\mu^2/E^2}{\mu^2/E^2}\right)-2\right],
\end{eqnarray}
where $E$ is the center of mass energy of the colliding parton pair.
In obtaining the second line of above equation, we have used
Eq.~(\ref{cross}). For quark-gluon plasma of massless quarks and
gluons at temperature $T$, we have $\langle p\rangle=3T$ and
$E\sim\sqrt{18}T$. The entropy density $s$ of the
quark-gluon plasma, which is modeled by quarks and antiquarks of
current masses in the AMPT, is simply
$s=(\epsilon+P)/T=4\epsilon/(3T)=96T^3/\pi^2$ if only up and down
quarks are considered.  The resulting specific viscosity is thus
\begin{eqnarray}\label{vis}
\eta_s/s\approx
\frac{3\pi}{40\alpha_s^2}\frac{1}{\left(9+\frac{\mu^2}{T^2}\right)\ln\left(\frac{18+\mu^2/T^2}{\mu^2/T^2}\right)-18}.
\end{eqnarray}

The initial temperature of heavy ion collisions in the AMPT model
can be estimated from the average energy density of mid-rapidity
partons at their average formation time, which is about $46.0$
GeV/fm$^{3}$ at LHC and $19.5$ GeV/fm$^{3}$ at RHIC. Using
$\epsilon=72T^4/\pi^2$ for the baryon-free quark and antiquark
matter, we obtain an initial temperature of about $468$ MeV at LHC
and about $378$ MeV at RHIC. The value of $\eta_s/s$ in the early
stage of the quark-gluon plasma formed in these collisions, when
most of the elliptic flow is generated, is thus about $0.273$ at LHC
and $0.085$ (for $\mu=1.8$ fm$^{-1}$) or 0.114 (for $\mu=2.3$ fm$^{-1}$) at RHIC.

We note that for fixed values of $\mu$ and $\alpha_s$, which is the
case in the AMPT model, the value of $\eta_s/s$ as given by
Eq.~(\ref{vis}) increases as the temperature of the quark-gluon
plasma decreases. In a more realistic description, the screening
mass depends on temperature, i.e., $\mu=gT$ with
$g=\sqrt{4\pi\alpha_s}$~\cite{Blazoit02}. In this case, the ratio
$\eta_s/s$ becomes independent of temperature. It is then of great
interest to extend the AMPT model to determine the screening mass
from the QCD coupling constant and the local temperature of the
partonic matter as in Ref.~\cite{Zhang10} and use the resulting
model to study if using the same QCD coupling constant $\alpha_s$ or
same specific viscosity $\eta_s/s$ in the AMPT model would describe
the elliptic flow in heavy ion collisions at both RHIC and LHC, as
in the findings of Refs.~\cite{Luz10,Lac10} based on the
hydrodynamic model.

\section{Conclusions}
\label{summary}

We have used the  AMPT model with string melting to study Pb-Pb
collisions at center of mass energy $\sqrt{s_{NN}}=2.76$ TeV. We
have found that the measured multiplicity density and elliptic flow
of charged particles at mid-pseudorapidity can be reasonably
described by the model if the parameters in the Lund fragmentation
function are taken to be those used in the default HIJING model and
that a smaller but more isotropic parton scattering cross
section than that used for heavy ion collisions at RHIC is used. As
at RHIC, the final-state partonic and hadronic scatterings were
found to be important as they would reduce the charged particle
multiplicity density at mid-pseudorapidity by about
$25\%$. The smaller parton cross section needed to
describe the measured elliptic flow at LHC than at RHIC has led to a
larger estimated $\eta_s/s$ in the quark-gluon plasma produced at
LHC than that at RHIC. However, this result may be due to the use of
constant screening mass in calculating the parton scattering cross
section. Taking into account the medium dependence of the screening
mass may reduce this difference. Furthermore, the transverse
momentum spectra and the centrality dependence of the multiplicity
and elliptic flow have been studied and they are roughly consistent
with the ALICE data. We have also made predictions for the
multiplicity distributions and elliptic flows of identified hadrons
such as protons, kaons, and pions. Comparisons of these predictions
with future experimental data will further enhance our understanding
of heavy ion collision dynamics at LHC and the properties of
produced quark-gluon plasma.

\begin{acknowledgments}
This work was supported in part by the U.S. National Science
Foundation under Grant No. PHY-0758115, the US Department of Energy
under Contract No. DE-FG02-10ER41682, and the Welch Foundation under
Grant No. A-1358.
\end{acknowledgments}

\end{document}